\documentstyle[prl,aps,twocolumn,epsf,amssymb,graphicx]{revtex}

\begin{document}

\title{Spin stripes in nanotubes}
\author{Alex Kleiner}
\address{Institute of Theoretical Physics\\  Chalmers
University of Technology and G\"{o}teborg University\\  
S-412 96 G\"{o}teborg, Sweden \\ }

\wideabs{ 
\maketitle
\begin{abstract}
It is shown here that electrons on the surface of a nanotube
in a perpendicular magnetic field undergo spin-chirality
 separation along the circumference.
Stripes  of spin-polarization propagate  along the tube,
with a spatial pattern that can be modulated by the electron filling.

\end{abstract}
\begin{figure}
      \begin{center}
 \includegraphics[width=3.35in]{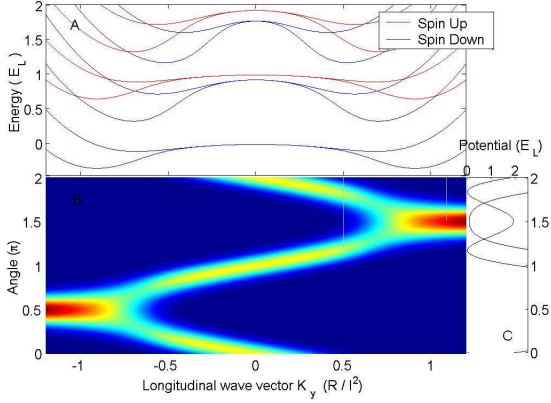} 
      \caption{\label{spectrum}Energy - Probability density  phase
        space.~~(A): Energy spectrum of eq. (\ref{small ham}),
        calculated numerically for $R=2.75 l$ 
        ($R=50\textrm{nm}$ and $B=2\textrm{T}$), $m^*=m_e$ and $g=2$.
        ~~(B): Spatial probability distribution  
        along the
        circumferential coordinate
        of the lowest band $(n=0)$ states in the spectrum.
        Here the electrons are well confined to the proximity of their
        potential minima. At $K_y=0$ the
        potential in (\ref{small ham}) is a double-well, one at each
        pole. As $|K_y|$ increases, the two wells move closer towards
        one of the equators, and when
        $|K_y|\ge R/l^2$ they merge to one well, at the equator.
        This point is illustrated in (C) where we zoom on the potentials
        of two states,
        $K_y=0.5R/l^2$ and $1.1R/l^2$ marked in (B) with white
        lines, having  a double-well and a single well potentials,
        respectively.
        }
      \end{center} 
\end{figure}

\pacs{73.63.Fg, 75.75.+a}}
The emerging field of spintronics opens a new
 paradigm
 of electronics based on the electron's spin rather than
 charge \cite{spintronics}.
 The realization of a spin circuit requires the
 generation, conduction and manipulation of spin currents.
 Recently, a spatial control of spin currents was demonstrated by
 engineering a material with a spatially varying g-factor \cite{g factor}. 
 An alternative route would require a magnetic field with spatial
 variation on the nanometer scale. Strong variations of the 
 field can be achieved along the circumference of a nanotube subjected to a
 uniform magnetic field, directed perpendicular to its axis. 
Here I suggest that a two dimensional
 electron gas (2DEG) rolled-up to a nanotube,
 may form spin-stripes propagating in alternating directions, as a
 function of the filling.
 This applies to the conduction electrons in fields satisfying  $l\lesssim R$, 
  where $l$ is the Landau length $\sqrt{\hbar/|eB|}$ and $R$ is the
 tube radius. 
 At magnetic fields of B$\lesssim 10$T for example, the radius
 should be 
 $R\gtrsim 8$nm. Such sizes frequently occur in  

 multi-wall carbon nanotubes (MWCNT) and in the new class of
  recently produced rolled-up heterostructures \cite{prinz rolled up}
 \cite{oliver 2001}. 
 The former, showed magneto-conductance fluctuations
 \cite{shon}\cite{korean}, with varying interpretations in connection
 to the 
 diffusive \cite{shon comment} or 
 ballistic \cite{korean reply}\cite{roche} nature of the MWCNT
 charge conductance.
 The spin conductance of a MWCNT however, was shown to be ballistic \cite{kazuhito 1999}
 over fairly large distances ($\gtrsim 130$nm). 
 On the other hand, the cylindrical heterostructures, made of silicon,
 silicon-germanium \cite{oliver 2001} or indium-gallium and
 indium-arsenic \cite{prinz rolled up}
 have the advantage of controlled radiuses that can easily satisfy  $l\lesssim R$, 
 they can be made clean and without the problem of unknown chirality
 and inter-shell coupling. 
 It was found numerically \cite{ajiki 93} that a cylindrical spinless two
 dimensional electron gas (2DEG) 
 under a perpendicular magnetic field, forms
 Landau level like states at the top and bottom and chiral states,
 similar to the edge states in the Hall bar, 
at the sides.\\
The magnetic field $B$ is taken here to be perpendicular to the surface
at the lines $x=0$ and
$x=\pi R$ hereafter the north and south
 `poles'. The `equators' are at $x=\pi R/2$ and $x=3\pi R/2$, and states located
 anywhere above or bellow the equators are called here `north' or `south' states.
 The vector potential on the surface of the tube is then  
$
\vec{A}= \left (0,RB\sin\frac{x}{R} \right ),
$
where $(x,y)$ are the circumferential and
axis directions of the tube, respectively. 
The Hamiltonian of a cylindrical 2DEG in this field is,
\begin{equation}\label{sch_ham}
H=-\frac{\hbar^2\partial_x^2}{2m^*}+\frac{\hbar^2}{2m^*}\left
  (-i\partial_y+\frac{eRB}{\hbar}\sin\frac{x}{R}\right )^2 +\mu
  g\bold{s\cdot B}
\end{equation}
where $m^*$ is the effective mass, $\mu$ is the Bohr magneton, $g$ is
the gyro-magnetic factor and $\textbf{s}$
is the spin operator.
The longitudinal wave vector and spin are conserved since the
Hamiltonian (\ref{sch_ham})  does not contain the $y$
coordinate nor other spin operators and so the operators are replaced by
their eigenvalues $K_y$ and $\pm g\mu B/2$.
The wave functions
for the spin-up and spin-down particles are now 
$
\psi_{\uparrow,\downarrow}=e^{iK_yy}\chi_{\uparrow,\downarrow}(x).
$ In units of $E_L/2$, where $E_L=eB\hbar/m^*$ is the Landau
level energy spacing,
eq. (\ref{sch_ham}) becomes the following one dimensional Hamiltonian, 
\begin{equation}\label{small ham}
H=-l^2\partial_x^2+\left
  ( K_yl+\frac{R}{l}\sin\frac{x}{R}\right )^2 \pm\frac{gm^*}{2m_e}
\end{equation}
The Hamiltonian  (\ref{small ham}) is a variant of Hill's
equation and can be easily  diagonalized numerically
 \cite{ajiki 93}.
We want to work in the regime where all the wave-functions are confined
in the circumferential direction.
The weakest confining potential in  (\ref{small ham}) is for $K_y=0$, which
is a double-well with minima at the poles. This potential gives, to a
linear order in $x$, Landau levels centered at the poles, with  
a spatial extension of  $l\sqrt{2n+1}$, where  $n=0,1,2\cdot\cdot$.
Thus, the potential is always confining if 
 $R\gtrsim l\sqrt{2n+1}$. 
The typical energy spectrum and probability distribution
 in this regime are shown in fig. (\ref{spectrum}).
  
\begin{figure}
      \begin{center}
 \includegraphics[width=3.35in]{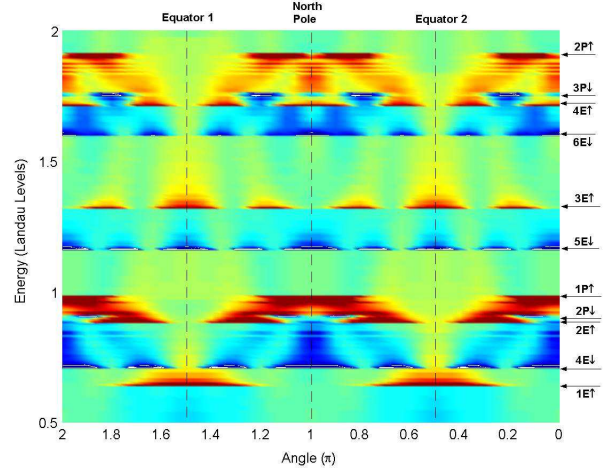} 
      \caption{\label{dos} Spin polarization distribution along the
        circumference vs. energy, with parameters as
               in fig. (\ref{spectrum}). 
               The pronounced polarization densities are mainly at energies in the proximity of
               singularities in the density-of-states of either
               spin-up or spin-down bands (red
               and blue, respectively). Each singular spin
               state  
               is either centered at a pole or around an equator.
               It is marked on the right with P or E
               respectively. e.g: the arrow labeled
               1P$\uparrow$ marks a singular state in the \emph{first}
               sub-band, centered at the \emph{Pole} 
               having a spin-\emph{up}. 
               } 
      \end{center} 
\end{figure}

 The eigenfunctions (see fig. \ref{spectrum}B) are confined
 in the circumferential direction to their potential minima,
 depending on $K_y$.
 Since the Hamiltonian (\ref{small ham})
 is symmetric under a simultaneous 
 sign inversion of
 $x$ and $K_y$, states with opposite $K_y$  are centered at
opposite sides of the circumference \cite{ajiki 93}, and states with $K_y=0$ are thus
centered at the poles. 
In the limit of a vanishing magnetic field, each band is four-fold
 degenerate, i.e: twice due to spin degeneracy and twice due to
 clock-wise and counter clock-wise propagating modes. The magnetic field
 removes the four degeneracies, as shown in fig. (\ref{spectrum}A),
 except at $K_y\approx 0$, where a two-fold degeneracy remains.
 Higher magnetic fields will not remove this degeneracy but rather
 increase it, since here, in the confinement regime, the potential for
 $K_y\approx 0$ has two deep and
 isolated potential wells at the two poles. Only as $K_y \rightarrow R/l^2 $ the
 two potential wells get close to each other across one of the
 equators 
 for their corresponding states to mix and remove the
 degeneracy. The total energy can be approximated analytically (see
 note \cite{note})
 for small $K_y$'s to give 
\begin{equation}\label{En}
 E=\hbar\omega
(n+\frac{1}{2})+3\lambda\left(\frac{\hbar}{2m\omega}\right)^2(2n^2+2n+1)\pm\Delta
E_n +2s
\end{equation}
where $\omega, \lambda$ and $\Delta E_n$ are functions of $K_y$, given
in the note \cite{note} and $m\equiv m_e$, having set for simplicity
$m_e=m^*$ and $g=2$ in Eq. (\ref{small ham}).
 The first two terms in Eq. ( \ref{En}) are the 
energies of a harmonic oscillator with an 
anharmonic correction of a  single-well potential $V(K_y)$ at 
 a minima of Eq. (\ref{small ham}). Since Eq. (\ref{small ham}) has
 two minimas for $|K_y|< R/l^2$, these terms alone would give a 
 two-fold degeneracy, without counting the spin. The third
 term largely removes this degeneracy by mixing the north and south
 states, and the last term is the Zeeman splitting, with
 $s=\pm\frac{1}{2}$.
 Since the conduction properties are determined only by electrons at 
 the Fermi-energy, 
 having the map between the energy-momentum-spin state and
 the spatial
 distribution of that state (fig. \ref{spectrum}), we can now find 
 the spatial distribution of the conduction electrons and their spins.
The spin polarization density at a given Fermi-energy E is defined as
 \\$P(E,x)=\left(P_{\uparrow}(E,x)-P_{\downarrow}(E,x)\right)/\left(P_{\uparrow}(E,x)+P_{\downarrow}(E,x)\right)$, where the spin-up
 or spin-down polarization $P_{\uparrow,\downarrow}(E,x)= \sum_{K_y} g(E,K_y)
 |\chi_{\uparrow,\downarrow}(E,K_y,x)|^2$ factors the probability densities
 with the corresponding density-of-states, summed over all states $K_y$
 at the energy E.
 Fig. (\ref{dos}) shows the energy dependent spatial
 spin-polarization. It is dominated by states
 at energies with a 
 divergent density of states
 $g(E)=\left(\frac{dE}{dK_y}\right)^{-1}$ at some proximity. As evident
 from fig. (\ref{spectrum}), these are either the Landau-like states
 at the poles having $K_y=0$, or states centered around the equators,
 to be reffered to as pole and equator singularities, respectively.

\begin{figure}
      \begin{center}
 \includegraphics[width=3.35in]{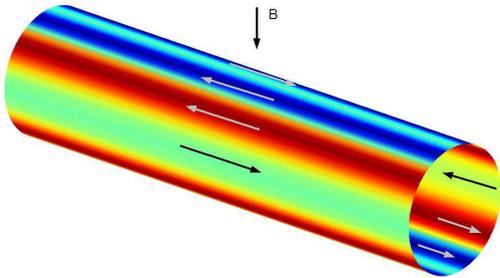} 
      \caption{\label{two tubes}Spin and chirality polarization.
        The Fermi-energy lies between the 2E$\uparrow$ and the 2P$\downarrow$
        singularities in fig. (\ref{dos}). 
        Red and blue colors represent, as
        in fig. (\ref{dos}), spin-up and spin-down polarizations, respectively.
        The
        arrows give the direction of propagation (chirality), taken
        from the sign of  $\frac{dE}{dK_y}$. The black or white color of some
         arrows is for visibility. 
        }
      \end{center} 
\end{figure}

 We can follow, for example, the spatial distribution of the equator 
singular states carrying
 spin-up. There are four such states in fig. (\ref{dos}), 
 marked at the right as
 $1E\uparrow$ to $4E\uparrow$, where the corresponding wave-functions around
 each equator have, one to four peaks, respectively .
 A similar
 observation can be made for the pole singularities (marked with P in
 fig. \ref{dos}),
 which are the Landau-like states. Their energy can be found analytically
 by simply
 setting $K_y=0$ in Eq. (\ref{En}), giving
\begin{equation} \label{landau}
E_n=E_L(n+\frac{1}{2}\pm\frac{1}{2}) - E_R(2n^2+2n+1),
\end{equation}
 where the first term is the usual Zeeman split Landau levels and the
 second term is the curvature correction due to the lateral
 energy, $E_R=\frac{\hbar^2}{8mR^2}$.\,\,\,\,  
 The spin-polarization $P(E,x)$ is
 dominated by the spin of the singular state with the closest energy
 to $E$. However, when the Fermi energy lies between the energies of singular
 states with opposite spins, such as between $2E\uparrow$ and
 $2P\downarrow$, or between $4E\uparrow$ and $3P\downarrow$ in
 fig. (\ref{dos}), there is
 a coexistence of spin-up and spin-down
 with different spatial distributions. This gives rise to the
 formation of spin-polarized stripes on the surface of the tube.
 Fig. (\ref{two tubes}) shows the spin stripes at 
 that energy, with the additional information
 on the chirality of these stripes.

 There is a 
 rather complex pattern of left moving and right moving
 spin `lanes'. The fact that spin distribution of the highest occupied
 electron states is entirely dependent on the filling, as shown in
 the polarization map (fig. \ref{dos}) suggests that a gate voltage may
 control the spatial distribution of spins. 
 Experimentally, 
 it appears feasible to observe
 the spin-stripes spatial structure by a 
 spin-polarized STM (SP-STM) \cite{heinze 2000}, at temperatures
 $kT\ll E_L$.
 The vertical structure of the polarization map (fig. \ref{dos}) may
 be observed by sweeping the gate voltage with the conventional
 two or four point contact set-up.
 For short tubes,
 $K_y$ in fig. (\ref{spectrum}) becomes discreet,
 with most of the allowed states fall at energies were the
 density-of-states diverges. In other words, the polarization map
 will converge to the discreet levels
 marked by the arrows on the right in fig. \ref{dos}. Only one third
 of these levels, the pole states (eq. \ref{landau}) can be considered as
 modified levels of the  `flat' quantum dot. The other levels are intrinsic
 to the tube.
 This could be supported by the experimental observation
 \cite{tans} that the spin dependent energy levels of a 
 short carbon nanotube `dot', do not follow the simple, flat, 
 quantum dot level filling. However, the filling pattern in \cite{tans}
 can not be 
 expected to follow the 
 polarization map (fig. \ref{dos}) since 
 it was not
 conducted in the confinement regime so that other reasons, such as
 electron-electron interactions, may have
 played a more important role, as suggested by the authors.
 The confinement condition for the experimental resolution of
 spin-stripes requires either large fields or large radiuses. 
  e.g: if the radius is in the range 
 $5\textrm{nm}<R<25\textrm{nm}$, at the lowest filling $n=0$, 
 the confinement condition $R\gtrsim l\sqrt{2n+1}$ 
 gives $B\gtrsim 20\textrm{T}$ and $B\gtrsim 1\textrm{T}$,
 where the higher field corresponds to the lower radius. These
 conditions, as already noted, can be easier achieved using the 
 cylindrical heterostructures. 
\\
In conclusion, it was shown that when a nanotube is subjected to a
 perpendicular magnetic field, under the specified conditions, there is
 a formation of spin-stripes on the surface of the tube with different
 propagation directions. The sensitivity of the spin pattern to the
 filling energy opens a potentially new way to generate and manipulate
 spin currents with a gate.
Finally, the stripe formation may be 
tested directly by the recently demonstrated \cite{heinze 2000} spin-polarized
scanning tunneling microscope (SP-STM).

\acknowledgments
I am indebted to Sebastian Eggert for many discussions and
 to Kim Jaeuk, Mikael Fogelstr\"om and Paata Kakashvili for helpful
comments on the manuscript.



\begin{thebibliography}{99}
\bibitem{spintronics}
S. A. Wolf et al., Science \textbf{294}, 1488 (2001) 
\bibitem{g factor}
G. Sallis et al., Nature \textbf{414}, 619 (2001)
\bibitem{prinz rolled up}
V. Ya. Prinz et al. Physica E \textbf{6}, 828 (2000)
\bibitem{oliver 2001}
Oliver G. Schmidt, Karl Eberl, Nature \textbf{410}, 168 (2001)
\bibitem{shon}
C. Sch\"onenberger et al. Appl. Phys. A \textbf{69}, 283 (1999)
\bibitem{korean}
J. O. Lee et al. Phys. Rev B \textbf{61}, R16362 (2000)
\bibitem{shon comment}
C. Sch\"onenberger and A. Bachtold, Phys. Rev. B \textbf{64}, 157401
(2001)
\bibitem{korean reply}
J. Kim et al., Phys. Rev. B \textbf{64}, 157402 (2001)
\bibitem{roche}
S. Roche and R. Saito, Phys. Rev. Lett. \textbf{87}, 246803 (2001)
\bibitem{kazuhito 1999}
K. Tsukagoshi, B. W. Alphenaar and H. Ago, Nature \textbf{401}, 572 (1999).
\bibitem{ajiki 93}
H. Ajiki and T. Ando, J. Phys. Soc. Jpn. \textbf{62}, 1255 (1993)
\bibitem{note}
For $R\gtrsim l\sqrt{2n+1}$ all states are confined to the proximity of their potential minima, which
are at $x_{min}=R\sin^{-1}\frac{|K_y|l^2}{R}$ and $\pi R-x_{min}$ for $-R/l^2\le K_y \le 0 $, and for states
 with $0\le K_y
\le R/l^2$ at $2\pi R-x_{min}$ and
$\pi R+x_{min}$.
The potential at a small distance $\epsilon$ from the minima can be
expanded as 
$V(K_y,\epsilon)=\frac{1}{2}m\omega^2\epsilon^2+\lambda\epsilon^4$, 
with $\omega=\frac{1}{R}\sqrt{\frac{E_l}{m}(\frac{R^2}{l^2}-K^2_y\,l^2)}$ and 
$\lambda=\frac{E_l}{24R^4}\left(7K^2_y\,l^2-4\frac{R^2}{l^2}\right)$. 
The potential at  $|K_y|\le R/l^2$ is a double-well,
symmetric about the equator, described by $V(K_y,\epsilon)$  at
the vicinity of its two minima $x_{min}$ and $\pi R- x_{min}$ .
 Denoting  $\psi_{1/2}$
 for the two single well wave functions, the total wave function of state
 $K_y$, to a zeroth order is $\Psi_{s/a}=\frac{1}{\surd 2}(\psi_1\pm\
 \psi_2)$, where s/a corresponds to the symmetric and antisymmetric
 product respectively. Due to the non-zero tunneling probability
 across the equator, the degeneracy of the symmetric and antisymmetric
 states is lifted.
 In the semi-classical approximation, the energy is split by
 $\frac{4\hbar^2}{m}\psi(x_{eq})\psi^\prime(x_{eq})$ where $\psi(x_{eq})$ and
 $\psi^\prime(x_{eq})$ are either one of the single well wave functions and
 their derivative, at the equator.
 Taking  $\psi(x_{eq})$ to be the
 eigenfunctions of the harmonic part of the energy, the first two
 energy splittings are $\Delta
 E_1=\frac{2\hbar^2}{m}\frac{1}{\sqrt{\pi}a^3}be^{-(b/a)^2}$ and $\Delta
 E_2=\frac{4\hbar^2}{m}\frac{1}{\sqrt{\pi}a^5}b^3e^{-(b/a)^2}$,
 where $a\equiv \sqrt{\frac{\hbar}{m\omega}}$ and
 $b=R(\pi/2-\sin^{-1}\frac{K_y \,l^2}{R})$. This result agrees with
 the numerics for $|K_y|\lesssim\frac{R}{2l^2}$.
\bibitem{heinze 2000}
Heinze S. et al., Science 288, 1805 (2000)
\bibitem{tans}
 Sander J. Tans et al., Nature \textbf{394}, 761 (1998)
\end{thebibliography}
\end{document}